\documentclass[aps,pra,floatfix,twocolumn,superscriptaddress,showpacs,longbibliography,nofootinbib]{revtex4-2}

\usepackage{graphicx}
\usepackage{amsmath}    
\usepackage{xcolor}
\usepackage{bibunits}
\usepackage{dcolumn}    
\usepackage{bm} 
\usepackage{amsmath}    
\usepackage{enumitem}
\usepackage{latexsym}
\usepackage{amsfonts}   
\usepackage{amssymb}
\usepackage{array}      
\usepackage{epsfig}
\usepackage{braket}
\usepackage{xcolor}
\usepackage{hyperref}

\newcommand{\ignore}[1]{}

\newcommand{\eq}{Eq.\,}
\newcommand{\eqs}{Eqs.\,}
\newcommand{\fig}{Fig.\,}

\newcommand{\cf} {cf.~}

\newcommand{\ie} {i.e.~}
\newcommand{\eg} {e.g.~}
\newcommand{\rref} {Ref.\,}
\newcommand{\rrefs} {Refs.\,}

\newcommand{\one}{1 \hspace{-1.0mm}  {\bf l}}
\newcommand{\SM}{\text{Supp.~Mater.}}
\newcommand{\PP}{\mathcal{P}}
\newcommand{\Res}[1]{\underset{#1}{\rm{Res}}}

\begin{document}

		\title{Dressed emitters as impurities}

\author{Luca Leonforte}
\affiliation{Universit$\grave{a}$  degli Studi di Palermo, Dipartimento di Fisica e Chimica -- Emilio Segr$\grave{e}$, via Archirafi 36, I-90123 Palermo, Italy}

\author{Davide Valenti}
\affiliation{Dipartimento di Fisica e Chimica ``Emilio Segrè", Group of Interdisciplinary Theoretical Physics, Università di Palermo, Viale delle Scienze, Ed.18, I-90128 Palermo, Italy}

\author{Bernardo Spagnolo}
\affiliation{Dipartimento di Fisica e Chimica ``Emilio Segrè", Group of Interdisciplinary Theoretical Physics, Università di Palermo, Viale delle Scienze, Ed.18, I-90128 Palermo, Italy}
\affiliation{Radiophysics Department, National Research Lobachevsky State University of Nizhni Novgorod, 23 Gagarin Avenue, Nizhni Novgorod 603950, Russia}
\affiliation{Istituto Nazionale di Fisica Nucleare, Sezione di Catania, Catania, Italy}

\author{Angelo Carollo}
\affiliation{Universit$\grave{a}$  degli Studi di Palermo, Dipartimento di Fisica e Chimica -- Emilio Segr$\grave{e}$, via Archirafi 36, I-90123 Palermo, Italy}
\affiliation{Radiophysics Department, National Research Lobachevsky State University of Nizhni Novgorod, 23 Gagarin Avenue, Nizhni Novgorod 603950, Russia}

\author{Francesco Ciccarello}
\affiliation{Universit$\grave{a}$  degli Studi di Palermo, Dipartimento di Fisica e Chimica -- Emilio Segr$\grave{e}$, via Archirafi 36, I-90123 Palermo, Italy}
\affiliation{NEST, Istituto Nanoscienze-CNR, Piazza S. Silvestro 12, 56127 Pisa, Italy}

\begin{abstract}
Dressed states forming when quantum emitters or atoms couple to a photonic bath underpin a number of phenomena and applications, in particular dispersive effective interactions occurring within photonic bandgaps.
	Here, we present a compact formulation of the resolvent-based theory for calculating atom-photon dressed states built on the idea that the atom behaves as an effective impurity. This establishes an explicit connection with the standard impurity problem in condensed matter. Moreover, it allows us to formulate and settle in a model-independent context a number of properties previously known only for specific models or not entirely formalized. The framework is next extended to the case of more than one emitter, which is used to derive a general expression of dissipationless effective Hamiltonians explicitly featuring the overlap of single-emitter dressed bound states.
\end{abstract}



\maketitle

\section*{Introduction}

Quantum emitters (``atoms") coupled to structured and/or low-dimensional photonic environments are an important paradigm of quantum optics and nanophotonics.
Important setups are photonic waveguides \cite{painter2019mirror}, a major focus of waveguide QED \cite{RoyRMP17,gu2017microwave,sheremet2021waveguide}, and enginereed photonic lattices  implemented in various ways such as coupled cavities/resonators, photonic crystals or optical lattices \cite{Sundaresan2019,painter205,scigliuzzo2021probing,KimblePNAS2016,Krinner2018}. 
Among others, a major appeal of such systems is the possibility of harnessing photon-mediated interactions between the emitters for implementing effective many-body Hamiltonians. Remarkably, for emitters energetically lying within photonic bandgaps, such effective second-order interactions can be dissipantionless. These are usually explained in terms of mediating {\it dressed} bound states (BSs) \cite{Douglas2015b, Gonzalez-Tudela2015,Calajo2016b, Gonzalez-Tudela2017b,Gonzalez-Tudela2018a,LeonfortePRL21}. In one such BSs, the atom is dressed by a photon exponentially localized in its vicinity. Overlapping single-emitter dressed BSs then results in an effective interatomic potential, somewhat similarly to the formation of molecules.

A quantum emitter coupled to a homogeneous photonic reservoir has interesting analogies with the textbook {\it impurity} problem in condensed matter \cite{Economou2006}. For instance, the reflection and transmission coefficients of a photon scattering off an atom in a waveguide are formally the same as those for an impurity described by an effective, energy-dependent, scattering potential \cite{ciccarelloPRAgate,PRA_SD}. 
Moreover, as first suggested in \rref\cite{Douglas2015b}, the aforementioned in-gap atom-photon BSs are quite reminiscent of bound states induced by an impurity inside lattice bandgaps \cite{Economou2006}. Recently, some of us identified a class of dressed states (dubbed ``VDS") whose photonic wavefunction is just the same that would arise if the atom were replaced by a vacancy (\ie an impurity described by a pointlike potential of infinite strength) \cite{LeonfortePRL21}. These states play a central role in topological quantum optics \cite{Bello2019,PainterPRX20,vega2021qubit} and, additionally, encompass a type of dressed bound states in the continuum (BIC) that are currently investigated in waveguide QED \cite{Longhi,TufarelliPRA13,GonzalezBallestroNJP13,RedchenkoPRA14,FacchiPRA16,Calajo2019,CornerPRR20,LonghiGiant2020}.

A natural question is to what extent the emitter-impurity analogy holds, in particular whether it can be formalized in a model-independent framework and applied to the calculation of both bound and unbound dressed states (the latter ones rule photon scattering processes). Mostly motivated by this question, this work introduces a novel formulation of the non-perturbative resolvent-based framework for studying atom-photon dressed states \cite{CohenAP, Lambropoulos2000a,Kofman1994}. We show that the resolvent operator (or Green function) can be expressed in a compact form structurally analogous to that arising in the standard impurity problem. This condenses the effect of atom-photon coupling in a single rank-one projector, thus allowing for a unified treatment of several kinds of dressed states (either bound and unbound). 
The framework is then extended to the case that more than one emitter is present and used to derive a general expression of dissipationless effective Hamiltonians (mentioned above), which explicitly connects the interatomic potential to overlapping single-atom dressed states independently of the specific photonic bath.

\section{Model and Hamiltonian}

	\begin{figure}
	\centering
	\includegraphics[width=\linewidth]{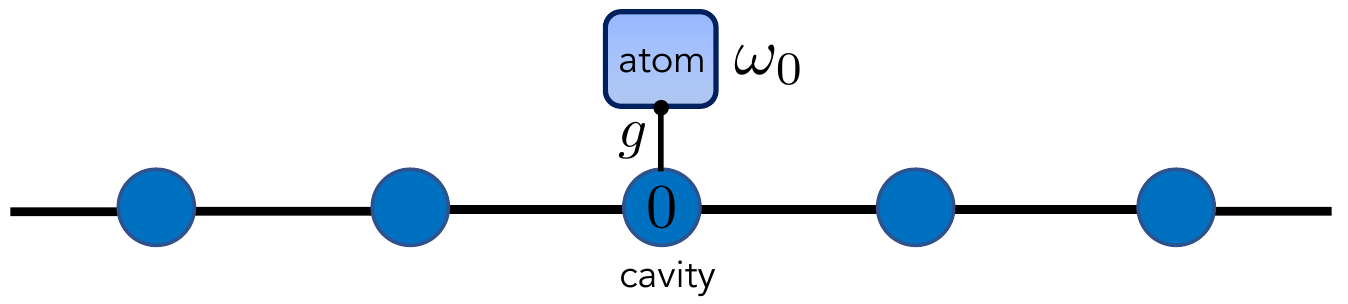}
	\caption{Schematic sketch of the considered model: a two-level quantum emitter (atom) coupled with strength $g$ to a photonic bath $B$ (field) modeled as a set of coupled cavities. The coupling is local in that the atom is directly coupled to only one cavity (labeled with 0). Here, $B$ is represented as a simple one-dimensional lattice (our framework applies to a generic bath).}
	\label{1at}
\end{figure}

We consider a two-level quantum emitter (``atom") coupled under the usual rotating-wave approximation to an unspecified photonic bath which is effectively modeled as a discrete set of $N$ coupled cavities (see skectch in \fig\ref{1at}). The Hamiltonian reads (we set $\hbar=1$ throughout the paper)
\begin{equation}
	\label{eq:genmode}
	H = H_0 + V
\end{equation}
with $H_0=H_e+H_B$ (unperturbed Hamiltonian), where
\begin{eqnarray}
	\label{eq:genmodel1}
	H_e & =&\omega_{0} \sigma_{+}\sigma_{-}\,, \\
	H_B&=&\sum_{x=1}^N \omega_x  b_{x}^\dag  b_x+\sum_{x\neq x'} \,J_{xx'}  b_{x}^\dag  b_{x'}\,,\label{HB}\\
	V & =& g \,( b_{0}^\dag  \sigma_{-} + b_{0}  \sigma_{+})\,\label{coupl}
\end{eqnarray}
with $\sigma_{-}=\sigma_{+}^\dag=|g\rangle\langle e|$ and $b_{x}$ the annihilation operators of the cavities fulfilling usual bosonic commutation rules ($\ket{g}$ and $\ket{e}$ are the atom's ground and excited states, respectively, and $\omega_0$ their energy separation). Here, $x=0$ (in the following sometimes referred to as the ``position" of the atom) labels the cavity directly coupled to the atom. 
Finally, $\omega_x$ is the bare frequency of cavity $x$, while $J_{x'x}=J_{xx'}^*$ denote the cavity-cavity hopping rates.

In all the remainder, we will be concerned solely with the single-excitation subspace spanned by $\{\ket{e}\ket{{\rm vac}},\,\{\ket{g}\ket{x}\}\}$ with $\ket{\rm vac}$ the field's vacuum state and $\ket{x}=b_x^\dag\ket{\rm vac}$ single-photon states. We conveniently introduce a light notation such that $\ket{e}\ket{{\rm vac}}\rightarrow \ket{e}$, $\ket{g}\ket{x}\rightarrow \ket{x}$, that is $\ket{e}$ ($\ket{x}$) is the state with one excitation on the atom ($x$th cavity).
Accordingly, in the single-excitation sector the Hamiltonian terms \eqref{eq:genmodel1}-\eqref{coupl} take the effective forms
\begin{eqnarray}
	H_e &=&\omega_{0}\ket{e}\!\bra{e}\,,\,\label{He2}\\
	H_B&=&\sum_{x=1}^N \omega_x  \ket{x}\!\bra{x}+\!\sum_{x\neq x'} \,J_{xx'}  \ket{x}\!\bra{x'}\,,\label{HB2}\\
	V & =& g \,( \ket{0} \!\bra{e}+\ket{e} \!\bra{0})\,\,.\label{coupl2}
\end{eqnarray}
The essential problem we are interested in is working out all the stationary states of the total Hamiltonian, \ie \,all single-photon dressed states.

\section{\label{sec-res}Resolvent}

The {resolvent} method \cite{Economou2006} is a powerful non-perturbative approach to compute eigenstates and eigenvalues of a Hamiltonian $H$, which is routinely used in many fields including quantum optics (see \eg \rrefs\cite{CohenAP,Lambropoulos2000a,Kofman1994,Compagno1999}). A few basics of the approach are briefly recalled next.

The  resolvent operator or Green function is defined as
\begin{equation}\label{Gz-gen}
	G(z)=\frac{1}{z-H}=\sum_n\frac{1}{z-E_n} \,\ket{E_n}\!\bra{E_n}\,\,
\end{equation}
where $z=\omega+i \omega'$ (with $\omega$, $\omega'$ real) 
runs over the entire complex plane and $H= \sum_n E_n \ket{E_n}\!\bra{E_n}$.
Energies $E_n$ and stationary states $\ket{E_n}$ (such that $H\ket{E_n}=E_n\ket{E_n}$) can be infered from the non-analyticity points of $G(z)$ on the $\omega$-axis (\ie the real axis). Specifically, stationary  {\it bound} states (BSs) are the residues of $G(z)$ around real poles (the poles being the corresponding energies). Instead, unbound eigenstates are associated with branch cuts of $G(z)$ on the real axis, each cut generally corresponding to an energy band of $H$ [at each point $\omega$ on a branch cut, $G(\omega+i\omega')$ jumps at $\omega'=0$]. Usually, the Hamiltonian is the sum of an unperturbed Hamiltonian $H_0$ and a perturbation $V$, $H=H_0+V$. In this case, if $\{\ket{ \varphi_k }\}$ are the unbound eigenstates of $H_0$ each with energy $E_{k}$, then the unbound stationary states of $H$ have the same spectrum and are worked out from these through the Lippmann–Schwinger equation as
\begin{equation}
	\ket{\Phi_k} = \ket{ \varphi_k } + {G}( E_k^+ ) V\ket{\varphi_k }\,,\label{LS}
\end{equation}
where we introduced the compact notation $h( \omega^+)  = \lim_{\delta \rightarrow 0^+} h ( \omega + i \delta )$. These thus fulfil $H	\ket{\Phi_k  } =E_{k}	\ket{\Phi_k  }$.
\\
\\
For our model in \eq\eqref{eq:genmode}, 
the bare atomic and field resolvents are respectively 
\begin{eqnarray}
	{G}_e ( z ) &=& \frac{1}{z-H_e}=\frac{\ket{e}\bra{e}}{z-\omega_{0}}\,,\label{Ge}\\
	{G}_B ( z ) &=& \frac{1}{z-H_B}=\sum_k \frac{\ket{k}\bra{k}}{ z - \omega_{k}} \,.\label{GBz}
\end{eqnarray}
Here, states $\{\ket{k}\}$ are the (single-photon) eigenstates of $H_B$ each corresponding to a normal mode of the field, \ie, $H_B \ket{k}=\omega_k \ket{k}$ with $\omega_k$ the associated normal frequency ($k$ coincides with the wavevector when $B$ is a lattice). Throughout the paper, we will assume that $G_B(z)$ {\it has no poles}, \ie $H_B$ has no bound eigenstates (as it is typically the case when $H_B$ describes a photonic lattice or waveguide). Indeed, the sum over $k$ in \eq\eqref{GBz} (and in similar expressions) must be intended as an integral weighted by the appropriate density of states.

\subsection{Resolvent in the impurity problem: review}

A longstanding topic in condensed matter and beyond is studying the effect of introducing an impurity into a lattice (although what follows does not require $B$ to be a lattice). 
The impurity is usually modeled as a contact potential of the form \cite{Economou2006}
\begin{figure}
	\centering
	\includegraphics[width=\linewidth]{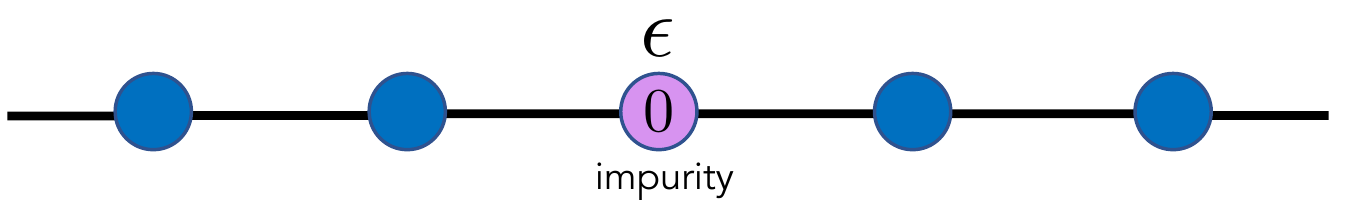}
	\caption{Impurity problem. A local static potential is added to the photonic bath, which is equivalent to a detuning $\epsilon$ changing the frequency of cavity 0 as $\omega_{x=0}\rightarrow \omega_{x=0}+\epsilon$.}
	\label{fig-imp}
\end{figure}
\begin{equation}
	V_{\rm imp} = \epsilon \, \ket{0} \bra{0}\,,\label{Hd}
\end{equation}
where $ \epsilon $ is the potential strength and $x=0$ the impurity position. 
When contextualized to our coupled-cavity lattice [\cf\eq\eqref{HB}], the impurity corresponds to an effective detuning of cavity $x=0$ [see \fig\ref{fig-imp}], which changes the field Hamiltonian as $H_B\rightarrow H_B+V_{\rm imp}$. Correspondingly, the field's resolvent $G_B(z)$ [\cf\eq\eqref{GBz}] turns into $G(z){=}(z-H_B-V_{\rm imp})^{-1}$. This can be worked out as \cite{Economou2006} 
\begin{equation}
	{G}( z ) =  {G}_B ( z ) + \frac{1}{ f(z)} \, \ket{\psi(z)}\!\langle \psi(z)| \,\label{G-eco}
\end{equation}
with
\begin{eqnarray}
\ket{\psi(z)}&=&{G}_B ( z ) \ket{0}\,,\label{psi1}\\
f(z)&=&\tfrac{1}{\epsilon}-  \braket{0| {G}_B ( z ) |0}\label{fz1}\,.
\end{eqnarray}
\eq\eqref{G-eco} states that the entire effect of the impurity is condensed in the rank-one projector featuring the $z$-dependent (unnormalized) state $\ket{\psi(z)}$. 

\subsection{Resolvent of the atom-field system}

Let us come back to our atom-field system and address now the total resolvent $G(z)$ corresponding to the total Hamiltonian $H$ [see \eq\eqref{eq:genmode}], i.e.\! \!when the atom-field coupling $V$ is included. As the atom can be seen as a sort of impurity although with internal degrees of freedom (``quantum impurity" \cite{Shen2007}), it is tempting to ask whether the total resolvent $G(z)=1/(z-H)$ with $H$ given by \eq\eqref{eq:genmode} can be expressed in a form structurally analogous to \eq\eqref{G-eco}. This is indeed possible, which can be shown through a transfer matrix method (see \SM). The result is\footnote{In \rref \cite{Lombardo2014b} an exact expression of $G(z)$ as a function of $G_B(z)$ was given, which is yet relatively involved and as such its physical interpretation not straightforward.}
\begin{equation}
	\label{eq:FullGreen}
	{G}(z) = {G}_B ( z ) + \frac{1}{ F ( z ) }\, \ket{ \Psi(z)} \bra{ \Psi(z)} \,,
\end{equation}
where, in place of \eq\eqref{psi1} and \eq\eqref{fz1}, we now have
\begin{eqnarray}
	\ket{ \Psi(z)}& =&\frac{1}{g}\,\ket{e} + \ket{\psi(z)}\,,\label{psiz}\\
	F(z) & =& \frac{1}{\epsilon(z)} -  \braket{0 | {G}_B ( z ) | 0}\label{fz}\,,
\end{eqnarray}
with 
\begin{equation}
	\epsilon(z)=\frac{g^2}{z-\omega_0}\,.\label{Vdz}
\end{equation}
Thus the fictitious impurity potential \eqref{Vdz} is $z$-dependent and scales as $(z-\omega_{0})^{-1}$. Interestingly, \eq\eqref{psiz} shows that $\ket{ \Psi(z)}$ can be seen as resulting from the hybridization of $\ket{\psi(z)}$, which is just the same as \eq\eqref{psi1}, with the atom. It is also worth observing that for $g=0$ the projector term in \eq\eqref{eq:FullGreen} correctly reduces to the bare atomic resolvent $G_e(z)$ [\cf\eq\eqref{Ge}].

Projecting the resolvent (see \eq\eqref{eq:FullGreen}) on the excitonic subspace, \ie \,on the state $\ket{e}$ (atom excited and field in vacuum), results in
\begin{equation}
	{\cal P}_eG(z){\cal P}_e=\frac{\ket{e}\!\bra{e}}{z-\omega_{0}-\Sigma(z)}\,\label{Pe}
\end{equation}
with the projector ${\cal P}_e$ defined as ${\cal P}_e=\ket{e}\!\bra{e}$. Here,
\begin{equation}
	\Sigma(z)=g^2\braket{0 | {G}_B ( z ) | 0}  \label{self-e}
\end{equation}
is the so called ``self-energy" \cite{CohenAP}. By comparing \eq\eqref{Pe} with \eq\eqref{Ge}, we retrieve the familiar picture \cite{Lambropoulos2000a,Kofman1994} according to which -- from the atom's viewpoint -- the effect of the interaction is to correct the atom's frequency $\omega_{0}$ with a $z$-dependent, generally complex, energy shift $\Sigma(z)$. 
\\
\\
Taking advantage of \eq\eqref{eq:FullGreen}, we can also straightforwardly project $G(z)$ on the field via the projector ${\cal P}_B=\sum_x \ket{x}\!\bra{x}$. Based on \eq\eqref{psiz}, we see that this is equivalent to replacing $\ket{\Psi(z)}$ with $\ket{\psi(z)}$ in \eq\eqref{eq:FullGreen}
\begin{equation}\label{PBG}
	{\cal P}_BG(z){\cal P}_B= {G}_B ( z ) + \frac{1}{ F ( z ) }\, \ket{ \psi(z)} \bra{ \psi(z)} 
\end{equation}
Comparing with \eqref{G-eco}, this shows that the field effectively behaves as if the atom were replaced by a fictitious impurity but with a $z$-dependent potential $\epsilon(z)$ [\cf\eqs\eqref{Hd}, \eqref{psi1} and \eqref{fz}]. This fact was already noted in \rref\cite{ciccarelloPRAgate} but for a specific dynamics and model. In contrast, \eq\eqref{PBG} shows that it is a general property, irrespective of the field Hamiltonian and interaction strength (so long as the rotating wave approximation holds). In a semantic analogy with the atom's self-energy, we will refer to $\epsilon(z)$ as the ``self-potential".
\\
\\
Thus, to sum up, while the atom acquires an effective self-energy $\Sigma (z)$, the field is subject to an impurity self-potential with strength $\epsilon(z)$.

\section{Dressed states}

In line with the previous section, we will first quickly review the scheme for calculating bound and unbound stationary states in the standard impurity problem and next consider dressed states in our atom-field system.

\subsection{Stationary states in the impurity problem}\label{ss-imp}

Applying the resolvent method (see first part of Section \ref{sec-res}) to the impurity problem, the energy $\omega_{\rm BS}$ of a stationary BS (if any) is a real pole of the resolvent (see \eq\eqref{G-eco}). Hence, it is a real solution of the pole equation \cite{Economou2006}
\begin{equation}	\label{eq:poleeqimp}
f(\omega)=0	 \,,
\end{equation}
\ie \!\! $\tfrac{1}{\epsilon}=\braket{0| G_B ( \omega )|0 }$
(recall that we assume throughout that $G_B(z)$ has no poles).
The associated stationary BS (density matrix form) is then just the residue of the resolvent \eq\eqref{G-eco} at $z=\omega_{\rm BS}$. The corresponding (unnormalized) BS ket is given by $\ket{\psi(z=\omega_{\rm BS})}$. Upon normalization, we get ${\cal N}\ket{\psi(\omega_{\rm BS})}$ with the normalization factor given by ${\cal N}=(\langle 0|{G}_B^2 ( \omega_{\rm BS} ) \ket{0} )^{-1/2}$.

The unbound stationary states $\{	\ket{\psi_k  }\}$ are instead obtained in terms of the unbound eigenstates $\{\ket{k}\}$ of $H_B$ [\cf\eq\eqref{GBz}] through the Lippmann-Schwinger equation \eqref{LS}. This generally yields (see \SM for details)
\begin{align}\label{psikk1}
	\ket{\psi_k  } = \left\{ \begin{aligned}
		& \,\ket{k } + \frac{\langle 0|k\rangle }{ f ( \omega_k ^+ ) }  \,\ket{ \psi( z=\omega_k^+ ) } && \text{ if } f(\omega_k) \neq 0\\
		&  \,\ket{k} && \text{ \rm if } f( \omega_k) = 0 
	\end{aligned} \right.
\end{align}
with $(H_0+V_{\rm imp}) \ket{\psi_k  }=\omega_{k}\ket{\psi_k  }$. 

We note that the latter case, namely $\omega_k$ such that $f( \omega_k) = 0$, occurs when there exists a solution of \eq\eqref{eq:poleeqimp} {\it within} a continuous band of $H_B$ (that is when, using the previous notation, $\omega_{\rm BS}\equiv\omega_{k}$ for some $\omega_{k}$). The corresponding eigenstate is then a BIC. \eq\eqref{psikk1} shows that all the unbound eigenstates of $H_B$ at this specific energy remain stationary states also in the presence of the impurity (thus in such a case $\omega_{\rm BS}$ will be at least twofold degenerate).
\\
\\
This highlights the role of states $\ket{\psi(z)}$ featured in the rank-one-projector term in the resolvent [\cf\eq\eqref{G-eco}], showing their close connection with stationary states. Indeed, the above can be summarized as follows: if there exists a real solution $\omega_{\rm BS}$ of \eq\eqref{eq:poleeqimp} then $\ket{\psi(z=\omega_{\rm BS})}$ (once normalized) is already a stationary state. This yields all the BSs. To obtain an unbound stationary state at energy $\omega_k$, we distinguish two cases: (i) $\omega_{k}$ does not fulfil \eq\eqref{eq:poleeqimp}, (ii) $\omega_{k}=\omega_{\rm BS}$ with $\omega_{\rm BS}$ a solution of \eq\eqref{eq:poleeqimp}. In case (i), the stationary state results from the superposition of $\ket{\psi(z=\omega_{k})}$, once weighted by $\langle 0|k\rangle/f(\omega_{k}^+)$, with the unperturbed unbound state $\ket{k}$. In case (ii), the stationary state simply coincides with the unperturbed state $\ket{k}$.

\subsection{Stationary states of the atom-field system}

Much like in the standard impurity problem where the BSs correspond to the real solutions of $f(\omega)=0$, the BSs of the atom-field total Hamiltonian \eq\eqref{eq:genmode} are obtained from the real solutions of $F(\omega)=0$ [recall \eqs\eqref{G-eco} and \eqref{eq:FullGreen}]. This is equivalent [\cf\eq\eqref{fz}] to the usual pole equation used in quantum optics in structured reservoirs \cite{Lambropoulos2000a,Kofman1994} 
\begin{equation}
	\label{eq:poleeq}
	\omega = \omega_0 + g^2\braket{0 | {G}_B ( \omega ) | 0}\,.
\end{equation}
Note that this equation (hence its solutions) differs from the impurity-problem analogue (see \eq\eqref{eq:poleeqimp}) ultimately due to the $z$-dependence of the impurity self-potential (see \eq\eqref{Vdz}). 

In formal analogy with the impurity problem, if $\omega_{\rm BS}$ is now a real solution of \eq\eqref{eq:poleeq}, then the corresponding dressed BS is just $\propto \ket{\Psi (z=\omega_{\rm BS})}$ [\cf\eq\eqref{psiz}]. Upon normalization, the dressed BS  can be arranged 
\begin{equation}
	\label{eq:wfbs}
\ket{\Psi_{\rm BS}} = {\cal N}\left( \ket{e} + g \,\ket{\psi ( \omega_{\rm BS} }\right)
\end{equation}
where we recall that $\ket{\psi (z)}=G_B(z)\ket{0}$ [\cf\eq\eqref{psi1}] and with the normalization factor given by
\begin{equation}
{\cal N} =\left(1+g^2 \langle 0|{G}^2_B ( \omega_{\rm BS}) \ket{0}\right)^{-\frac{1}{2}}\,\,.\label{norm}
\end{equation}
\\
Again in formal analogy with the impurity problem [see \eqs\eqref{psikk1}], unbound dressed states are found from the unperturbed unbound states $\{\ket{k}\}$ as (see \SM)
\begin{align}\label{psik}
\ket{\Psi_k  } =\left\{\begin{array}{c}
 \ket{k } +  \frac{\braket{ 0 | k}}{ F ( \omega_k^+  )}  \,\ket{ \Psi( z=\omega_k^+ ) } \,\,\, \text{ if } F(\omega_k) \neq 0\\
\ket{k}\qquad\qquad\qquad\qquad\,\,\,\,\,\, \,\,\,\,\text{ \rm if } F( \omega_k) = 0 
\end{array}	\right.\,.
\end{align}
\\
\\
To conclude this section, note that, like its impurity analogue (see end of Section \ref{ss-imp}), the state $\ket{\Psi(z)}$ in \eq\eqref{eq:FullGreen} is strictly connected with stationary states. A dressed BS (if any) is given by $\ket{\Psi(z=\omega_{\rm BS})}$ (to be normalized) with $\omega_{\rm BS}$ being a real solution of $F(\omega)=0$ [\ie of \eq\eqref{eq:poleeq}]. 
An unbound dressed state at energy $\omega_k$ not fulfilling \eq\eqref{eq:poleeq} instead results from the superposition of $\ket{\Psi(z=\omega_{k}^+)}$, once weighted by ${\braket{0 | k}}/{ F ( \omega_k^+  )} $, with the unperturbed unbound state $\ket{k}$. If instead $\omega_k=\omega_{\rm BS}$, \ie it is a solution of \eq\eqref{eq:poleeq}, the corresponding unperturbed $\ket{k}$ is a stationary state even in the presence of the atom which adds to the dressed BS $\propto\ket{\Psi(z=\omega_{\rm BS})}$\footnote{To our knowledge, this is the first general formulation of this property, which was discussed in \rref\cite{Longhi} but for a specific model.}. Note that the last property immediately entails the known impossibility to populate dressed BICs via single-photon scattering \cite{Calajo2019}: a photon at frequency $\omega_{0}$ sent from far away will just not see the atom.

\subsection{Vacancy-like dressed states}

An important special case occurs when there exists a dressed BS having the same energy as the atom, \ie $\omega=\omega_{0}$. If this is a BS then (as previously discussed) $F(\omega_{0})=0$, entailing 
\begin{equation}
	\langle 0|G_B(\omega_{0} ) |0\rangle=0\label{G00eq}
\end{equation}
[\cf\eqs\eqref{fz} or \eqref{eq:poleeq}]. This means that the photonic component $\ket{\psi(\omega_{0})}$ is just the all-photonic BS that would occur (at the same energy $\omega_{0}$) if the atom-photon interaction $V$ were replaced by $V_{\rm imp}$ with $\epsilon\rightarrow\infty$, \ie a vacancy on the cavity $x=0$ [recall \eqs\eqref{Hd}, \eqref{psi1} and \eqref{fz1}]. Also, \eq\eqref{G00eq} states that $\ket{\psi(\omega_{0})}=G_B(\omega_{0})\ket{0}$ has a node at $x=0$ (as it must be due to the infinite potential strength). If instead the dressed state at energy $\omega_{0}$ is unbound, then $\omega_0$ lies within some band and $F(\omega_{0})\neq 0$.
Using \eq\eqref{psik}, it can be expressed as
\begin{equation}
	\ket{\Psi_{k_0}} = \ket{k_0} - \frac{ \braket{0 | k_0}}{ \braket{0 | G_B ( \omega_0) | 0}} \ket{ \Psi ( \omega_0)},
\end{equation}
where $k_0$ is defined such that $\omega_{k_0} = \omega_0$. Projecting both sides on $\ket{0}$ and using again $\ket{\psi(\omega_{0})}=G_B(\omega_{0})\ket{0}$ [\cf\eq\eqref{psiz}] yield $\langle 0 \ket{ \Psi _{k_0}}=0$. Thus the dressed state also in this case has a node on $\ket{0}$.

This class of stationary states occurring at energy $\omega_{0}$, named ``vacancy-like dressed states" (VDSs), were introduced and studied in \rref\cite{LeonfortePRL21} without using the resolvent method. The above discussion shows that their properties are straightforwardly retrieved in our resolvent-based framework. More importantly, it clarifies the peculiarity of VDSs against all the other dressed states as follows. A dressed state can always be associated with an effective impurity seen by the field: VDS are the subset of dressed states whose corresponding impurity has an infinite potential strength (namely it reduces to a vacancy).

\section{\label{sec-2at}More than one emitter}

	\begin{figure}
	\centering
	\includegraphics[width=\linewidth]{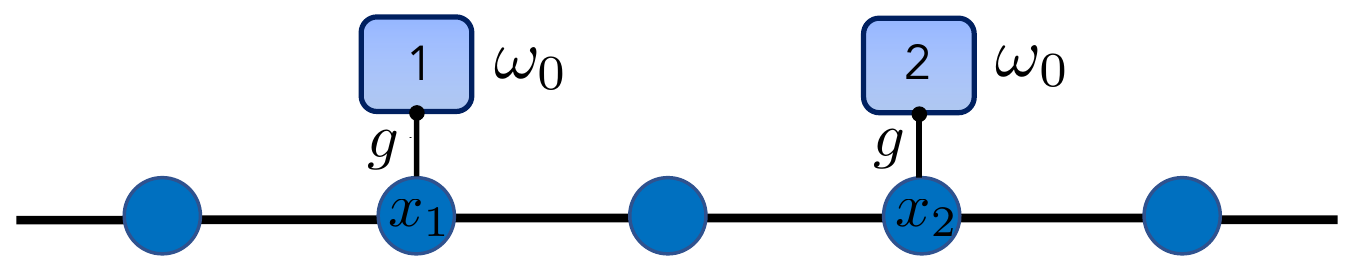}
	\caption{Schematics of two atoms, 1 and 2, coupled to a photonic bath (in this case skecthed as a simple lattice). Each emitter has frequency $\omega_0$ and is coupled to the cavity $x_i$ with strength $g$.}
	\label{2at}
\end{figure}
It is natural to ask how the one-emitter framework developed so far generalizes when another emitter is present. We thus consider next two atoms, labeled 1 and 2, each coupled to the photonic bath at site $x_i$ (for $i=1,2$) with strength $g$, see sketch in \fig\ref{2at}. Accordingly, \eq\eqref{He2} is now replaced by $H_e =\omega_{0}\sum_{i=1}^2\ket{e_i}\!\bra{e_i}$ (with $\omega_0$ the frequency of each atom), while \eq\eqref{coupl2} now turns into $V=  g\ket{x_1} \!\bra{e_1}+g\ket{x_2} \!\bra{e_2}+{\rm H.c.}$ (we consider equal couplings for the sake of simplicity).

The total resolvent in this case can be arranged as (see \SM\!\! )
\begin{align}
	{G} ( z ) = { G }_B ( z ) \!+\! \sum_{ij=1}^2 ({\bf F}^{-1})_{ij} \,\ket{ \Psi_i} \bra{ \Psi_j},\label{G2at}
\end{align}
where we defined the $z$-dependent states $\ket{\Psi_i(z)}$ and the $z$-dependent matrix ${\bf F}(z)$ as [their $z$-dependence is implicit in \eq\eqref{G2at}]
\begin{align}
	\ket{\Psi_i(z)} & = \frac{1}{g} \, \ket{ e_i } +   G_B ( z ) \ket{ x_i}\,, \label{Psii}\\
	F_{ij} ( z ) & =\frac{1}{\epsilon ( z ) }\delta_{ij} - \braket{ x_i | {G}_B ( z ) | x_j},\label{F2}  
\end{align}
with $\braket{x_j | \Psi_i}=\braket{\Psi_j|x_i}=\braket{ x_j | G_B ( z ) | x_i}$ and $\epsilon ( z ) $ as in \eq\eqref{Vdz}. Notice that, for $z \notin \mathbb R$, generally $F_{12}\neq F_{21}^*$, and that $\ket{\Psi_i(z)}$ and $F_{ii}(z)$ coincide with \eqs\eqref{psiz} and \eqref{fz}, respectively, when $\ket{0}$ is replaced by $ \ket{x_i}$.

This exact expression already points towards an effective interaction mediated by one-emitter dressed states. Indeed, the resolvent $G(z)$ reduces to $G_1(z)+G_2(z)$ (isolated emitters) when ${\bf F}(z)$ is diagonal, \ie $F_{12}=F_{21}=0$: this can only occur when $\ket{\Psi_i(z)}$ has a node on $x_{j\neq i}$ [\cf\eq\eqref{F2}]. This shows that the crosstalk between the emitters is mediated by $\ket{\Psi_i(z)}$ (via its photonic component $\ket{\psi_i(z)}$).

Each BS energy $\omega_{\rm BS}$ fulfils the pole equation ${\det {\bf F}(\omega_{\rm BS})}{=}0$, whose solutions are implicitly given by
\begin{equation}
	\omega_{\rm BS}^\pm = {\tilde \omega}_0(\omega_{\rm BS}^\pm) \pm \delta(\omega_{\rm BS}^\pm)\label{omegaBS}
\end{equation}
where the function ${\tilde \omega}_0(z)$ and $\delta(z)$ (here calculated at $z=\omega_{\rm BS}^\pm$) are defined as
\begin{align}
	{\tilde \omega}_0(z)  &= \omega_0 {+} \tfrac{g^2}{2}\sum_{i=1,2} \braket{ x_i| {G}_B(z) | x_i},\label{omega0}\\
	\delta(z)  &= \sqrt{g^4 F_{12}   F_{21}+{\cal A}^2},\label{delta}
\end{align}
with ${\cal A}(z)  = \tfrac{g^2}{2} \left(F_{22}-  F_{11}  \right)$ 
[the dependence on $z$ of $F_{ij}$ and ${\cal A}$ is implicit in \eq\eqref{delta}].

The residues of $G(z)$, $\ket{\Psi_{\rm BS}^\pm}\! \bra{\Psi_{\rm BS}^\pm}$, on the two poles $\omega_{\rm BS}^\pm$ can be calculated directly although their expression is too lengthy to be reported here (see \SM). The states $\ket{\Psi_{\rm BS}^\pm}$ are especially relevant when the emitters are far-detuned from all unbound photonic modes. Under weak-coupling condition, this guarantees that $\omega_{\rm BS}^\pm$ will also be far-detuned from the unbound photonic modes and close to $\omega_{0}$. Correspondingly, the emitters bare states $\ket{e_i}$ will have vanishing overlap with the unbound dressed states and lie (up to second order in $g$) in the eigenspace spanned by $\ket{\Psi_{\rm BS}^\pm}$. Thus, the atomic dynamics is described by the effective Hamiltonian $H_{\rm eff} \simeq  \sum_{i=\pm} \omega_{\rm BS}^{i} \ket{\Psi_{\rm BS}^{i}}\bra{\Psi_{\rm BS}^{i}}$. In this regime, we can derive the approximate explicit expressions for $\omega_{\rm BS}^\pm$ and $\ket{\Psi_{\rm BS}^\pm}$, which replaced in $H_{\rm eff}$ provide the following effective Hamiltonian (see \SM)
\begin{equation}
	H_{\rm eff} = H_{\rm s}+H_{\rm a}  \label{Heff1}
\end{equation}
with 
\begin{align}
	H_{\rm s} &= \lambda_{\rm s}\Big(\sum_{i=1,2}	\omega_{\rm BS}^{(i)} \ket{ \Psi_i}\! \bra{ \Psi_i} - (g^2 F_{12}\ket{ \Psi_1} \!\bra{ \Psi_2}+{\rm H.c.}) \Big),\\
	H_{\rm a} &= \lambda_{\rm a}\Big(\sum_{i=1,2}	\Omega_i \ket{ \Psi_i}\! \bra{ \Psi_i} - (g^2 F_{12}\ket{ \Psi_1} \!\bra{ \Psi_2}+{\rm H.c.}) \Big)\,,
\end{align}
where 
\begin{align}
 \omega_{{ \rm BS}}^{(i)}&=\omega_0 + g^2\braket{x_i|G_B(\omega_0)|x_i}\qquad (i=1,2)\\
 \lambda_{\rm s}&=-\frac{2 \delta^2 \,\left( \braket{ \Psi_1 |\Psi_1}  +  \braket{ \Psi_2| \Psi_2} \right)}{\beta^+\beta^-} \,,\\
 \lambda_{\rm a}&=\frac{2  {\omega}_{0}{\cal A} \,\left( \braket{ \Psi_1 | \Psi_1}  -  \braket{ \Psi_2 | \Psi_2} \right)}{\beta^+\beta^-}\,,\\
 \Omega_1&=	\frac{\delta^2}{{\omega}_{0}} + {\cal A} \,,\,\,\, \Omega_2= \frac{\delta^2}{{\omega}_{0}} - {\cal A} \,,\\
 \beta^\pm &= {\cal A} \left( \braket{ \Psi_1 | \Psi_1} \! {-}  \!\braket{ \Psi_2 | \Psi_2} \right)\pm\delta ( \braket{ \Psi_2 | \Psi_2} \!{+} \!\braket{ \Psi_1 | \Psi_1} )\,.
\end{align}
Here, all $z$-dependent quantities such as $\ket{\Psi_i}$, $F_{ij}$, $\omega_{\rm BS}^{(i)}$, $\mathcal{A}$ and $\delta$ [\cf\eqs~\eqref{Psii}-\eqref{delta}] are calculated at $z=\omega_0$. In particular, $\ket{\Psi_i}$ and $\omega_{\rm BS}^{(i)}$ are the (unnormalized) dressed BS of atom $i$ and the corresponding energy in the absence of the other emitter, respectively.

\eq\eqref{Heff1} expresses the effective Hamiltonian explicitly in terms of overlapping one-atom dressed BSs. A similar task was carried out in \rref\cite{Calajo2016b} yet for the specific model of a simple homogeneous photonic lattice. 

Notably, terms $H_{\rm s}$ and $H_{\rm a}$ are interpreted as follows. When the two atoms are located in equivalent positions then $\langle  \Psi_1|\Psi_1\rangle=\langle  \Psi_2|\Psi_2\rangle $ and thus $H_{\rm a}=0$. Otherwise, in general $\langle  \Psi_1|\Psi_1\rangle\neq \langle  \Psi_2|\Psi_2\rangle$ so that both $H_{\rm s}$ and $H_{\rm a} $ contribute to $H_{\rm eff}$. 
Thus the term $H_{\rm a}$ describes the effect of inequivalent emitters' positions. A noteworthy example is a translationally-invariant bath featuring a unit cell with more than one site. In this case it is known that changing from equivalent to inequivalent positions (or viceversa) can drammatically affect the effective coupling strength (which may even vanish) \cite{Bello2019,LeonfortePRL21}.

We conclude by noting that for equivalent positions we get the particularly compact form
\begin{align}
	H_{\rm eff} =H_s=\! \sum_{i=1,2}\!	\omega_{\rm BS}^{(i)} \ket{ \tilde\Psi_i}\! \bra{ \tilde\Psi_i}- \!(g^2F_{12}\ket{ \tilde\Psi_1} \!\bra{\tilde \Psi_2}+{\rm H.c.})
\end{align}
where $\ket{\tilde{\Psi}_i}={\braket{\Psi_i | \Psi_i}}^{-1/2}\ket{\Psi_i}$ is a normalized dressed BS (due to the hypothesis of equivalent positions, $\braket{\Psi_1 | \Psi_1} = \braket{\Psi_2 | \Psi_2}$).
\\
\\
The above arguments can be generalised (see \SM) to the case of $M$ emitters. Indeed, Eqs.~\eqref{G2at},~\eqref{Psii} and~\eqref{F2} are naturally generalized, yielding in the weak coupling regime the effective Hamiltonian
\begin{align}\label{eq:HeffM}
	H_{\rm eff} & =\! \sum_{i=1}^{M} \! \omega_{{ \rm BS}}^{(i)} \ket{\tilde \Psi_i}\bra{\tilde \Psi_i} - g^2 \sum_{i \neq j} F_{ij} \ket{\tilde \Psi_i} \bra{\tilde\Psi_j}\,
\end{align}
with $x_i$ the location of the $i$th atom and where $\ket{\tilde{\Psi}_i}$ and $\omega_{{ \rm BS}}^{(i)}$ are defined exactly as in the $M=2$ case.

\section{Conclusions}

To sum up, we considered a general model of quantum emitter coupled to an unspecified photonic bath under the rotating wave approximation. Inspired by analogies between an atom and a standard impurity, we have shown that the resolvent operator used in the non-perturbative description of atom-photon interactions can be re-arranged in a compact form so as to make it structurally analogous to that occurring in the textbook impurity problem. This complements the usual picture according to which the atom acquires a self-energy, showing that the field sees the atom as a fictitious impurity with an associated self-potential. As a hallmark, the presence of the atom in the resolvent is fully captured by a rank-one projector term appearing in the resolvent. This in turn features a dressed-state function $\ket{\Psi(z)}$ (defined on the complex plane) which in a sense encompasses already the stationary states (especially BSs). An extension of the framework to the case of more than emitter was carried out, which allows for a natural derivation of dispersive Hamiltonians explicitly in terms of overlapping one-atom dressed states.

This work settles the atom-photon resolvent formalism in a form arguably easier to handle. This can be beneficial for instance in view of generalizations to the topical paradigm of giant atoms \cite{Kockum5years}, \ie emitters non-locally coupled to the bath and as such more complicated to describe \cite{BurilloPRA20,KockumPRLTopo21,vega2021qubit}. Moreover, the connection with the familiar impurity problem makes the atom-photon resolvent apparatus physically more intuitive. We took advantage from the last circumstance in order to highlight some general properties having an impurity-problem counterpart (\eg the coexistence of dressed BICs with unperturbed photonic states).

\clearpage

\appendix



\onecolumngrid	

\section{Proof of Eqs.~(\ref{eq:FullGreen}) and~(\ref{G2at})}\label{app-TM}
We will derive Eqs.~(\ref{eq:FullGreen}) and~(\ref{G2at}) as a special case of a general formula for $M$ emitters coupled to the field. For $M$ emitters the Hamiltonian terms of section~\ref{sec-2at} generalise to \ref{eq:FullGreen}
\begin{eqnarray}
	H_e =\omega_{0}\sum_{i=1}^M\ket{e_i}\!\bra{e_i}\,,\,\label{He3}\qquad
	V = g \,\sum_{i=1}^M( \ket{x_i} \!\bra{e_i}+\ket{e_i} \!\bra{x_i})\,\,.\label{coupl3}
\end{eqnarray} 
with obvious meaning of the symbols, and $H_B$ remains as in Eq.~\eqref{HB2}.
In standard Green function theory \cite{Economou2006}, for a given a Hamiltonian $H=H_0+V$ and associated resolvent $G(z)=(z-H)^{-1}$, the T-matrix is defined as
\begin{align}
	T(z) & 
	\label{eq:transfMatrix} = V + T ( z) G_0 ( z) V \,.
\end{align}
where $G_0(z)=(z-H_0)^{-1}$ is the resolvent of the unperturbed Hamiltonian $H_0=H_e+H_B$. The resolvent can be expressed in terms of $T(z)$ as 
\begin{align}
	G(z)  = G_0 ( z ) + G_0 ( z )T(z) G_0 ( z ) \,.\label{GT}
\end{align}
Using the expression~\eqref{eq:transfMatrix} recursively yields a formal expression for $T(z)$ in series
\begin{equation}
	T ( z ) = \sum_{ k = 1}^{+\infty} T^k ( z )\,\,\,{\rm with}\,\,\,\,T_k ( z )=V \,( G_0 ( z ) V )^{k-1}=T_k ( z )=V \,( \PP G_0 ( z )\PP V )^{k-1}\,.
\end{equation}
where $\PP$ is the projector on the support of $V$. Since $V$ has only support on $\{\{\ket{x_i}\}_{i=1}^M,\ket{e_i}_{i=1}^M\}$, $\mathcal{P}$ is the projector operator onto this subspace. Notice that in the single particle sector $G_0(z)=G_e(z)+G_B(z)$, with $G_e(z)=(z-H_e)^{-1}$ and $G_B(z)=(z-H_B)^{-1}$. Thus, in the ordered basis $\{\ket{x_1},\dots,\ket{x_M}, \ket{e_1},\dots,\ket{e_M}\}$, 
the relevant operator assume the following block-matrix structures
\begin{align}\label{Matrice}
	\PP G_0(z)\PP\sim \begin{pmatrix}
		\bm{\gamma}^B(z)&0\\
		0&\bm{\gamma}^e(z)
	\end{pmatrix}\qquad\textrm{ and }\qquad
	V\sim g\, \begin{pmatrix}
		0&\one_M\\
		\one_M&0
	\end{pmatrix}\,,
\end{align}
where $\sim$ is a shorthand notation for matrix representation in the mentioned basis, $\one_M$ is the $M\times M$ identity matrix while $\bm{\gamma}^B(z)$ and $\bm{\gamma}^e(z)$ are $M\times M$ matrices with elements $\gamma^B_{ij}(z)=\braket{x_i|G_B(z)|x_j}$ and $\gamma^e_{ij}(z)=\braket{x_i|G_e(z)|x_j}=\gamma_e(z)\delta_{ij}$, respectively where $\gamma_e(z)=(z-\omega_0)^{-1}$.
Thanks to the simplicity of $V$, one can get explicit expressions for odd and even terms of $T(z)$ as
\begin{align}
	T_{2 k -1 } ( z ) \sim \,\, g &\left[ g^2\begin{pmatrix}\gamma_e ( z )\bm{\gamma}^B ( z )&0\\0&\gamma_e ( z )\bm{\gamma}^B ( z )\end{pmatrix} \right]^{k - 1} \cdot \begin{pmatrix}
		0&\one_M\\
		\one_M&0
	\end{pmatrix} \,,\nonumber\\
	T_{2 k } ( z ) \sim \,\, g^2 &\left[ g^2\begin{pmatrix}\gamma_e ( z )\bm{\gamma}^B ( z )&0\\0&\gamma_e ( z )\bm{\gamma}^B ( z )\end{pmatrix} \right]^{k - 1} \cdot \begin{pmatrix}
		\bm{\gamma}^e(z)&0\\
		0&\bm{\gamma}^B(z)
	\end{pmatrix} \,. \nonumber
\end{align}
We thus end up with
\begin{align}
	T ( z )  &\sim 
	g\begin{pmatrix}
		\bm{h}(z)&0\\0&\bm{h}(z)
	\end{pmatrix}
	\cdot \begin{pmatrix}
		g\bm{\gamma}^e(z)&\one_M\\
		\one_M&g\, \bm{\gamma}^B(z)
	\end{pmatrix}\,,\nonumber
\end{align}
where $\bm{h}(z)$ is a geometric series of $M\times M$ matrices, which for sufficiently weak coupling (\ie $|| g^2\bm{\gamma}^e(z)\cdot \bm{\gamma}^B ( z ) || {<}1$) converges, yielding
\begin{equation}
	\bm{h}(z)=\sum_{ k = 0 } \,\left[ g^2 \gamma_e ( z )\bm{\gamma}^B ( z )\right]^{k} \,=\frac{1}{ \one_M -  g^2 \gamma_e ( z )\bm{\gamma}^B ( z ) }.
\end{equation}
Plugging this in \eq\eqref{GT} we get the full Green function in the following form (where the repeated index summation convention is used and the $z$ dependences are  omitted),
\begin{align}
	\!\!	G ( z )  &= G_B+G_e + g^2 \gamma_e G_0\ket{x_i}  h_{ij}  \bra{x_j}G_0+ g G_0\ket{x_i}  h_{ij} \bra{e_j}G_0 + g G_0\ket{e_i}  h_{ij} \bra{x_j}G_0+g^2 G_0\ket{e_i}  h_{ij}\gamma_{jk}^B \bra{e_k}G_0\nonumber\\\label{GGG}
	&= G_B+\gamma_e\ket{e_i}\bra{e_i} + \gamma_e h_{ij} \left\{ g^2 G_B\ket{x_i}  \bra{x_j}G_B+ g G_B\ket{x_i} \bra{e_j} + g\ket{e_i}  \bra{x_j}G_0+g^2 \gamma_e\ket{e_i}\gamma_{jk}^B \bra{e_k} \right\}\nonumber\\
	&= G_B+ g^2\gamma_e \,h_{ij} \ket{\Psi_i}\bra{\Psi_j}
\end{align}
with $h_{ij}$ denoting the elements of the matrix $\bm{h}(z)$. We recall that $\ket{\Psi_i(z)}=\tfrac{1}{g}\ket{e_i}+G_B(z)\ket{x_i}$.
Finally, we obtain the expression
\begin{align}\label{GM}
	G(z)=G_B(z)+\sum_{ij} \{\bm{F}^{-1}(z)\}_{ij=1}^M\ket{\Psi_i(z)}\bra{\Psi_j(z)}
\end{align}
where $\bm{F}(z)=\frac{\one_M}{g^2\gamma_e(z)}-\bm{\gamma}^B(z)$, whose elements are $F_{ij}(z)=\frac{z-\omega_0}{g^2}\delta_{ij}-\braket{x_i|G_B(z)|x_j}$\,. For M=1, Eq.~\eqref{GM} reduces to Eq.~\eqref{eq:FullGreen} upon identifying the matrix $\bm{F}(z)$ with the scalar $F(z)$. For $M=2$, we recover Eq.~\eqref{G2at}. 

\section{Impurity problem}\label{sec-imp}

The total Hamiltonian reads (see main text) $H=H_B+V_{\rm imp}$ with $V_{\rm imp}= \epsilon\, \ket{0}\! \bra{0}$. One can show \cite{Economou2006} that the transfer matrix in this case is
\begin{equation}
	T ( z )  = \ket{0} \frac{\epsilon}{ 1 - \epsilon \braket{ 0 | {G}_B ( z ) | 0 } } \bra{0}
\end{equation}
if $|\epsilon \braket{ 0 | {G}_B ( z ) | 0} | < 1$.
The total Green function for this model is then
\begin{equation}
	{G} ( z ) = {G}_B ( z ) +  {G}_B ( z ) T ( z ) {G}_B ( z ) ,
\end{equation}
which coincides with \eq\eqref{G-eco} in the main text.

The only possible BS (if any) is the solution of the pole equation $\gamma_0 ( \omega_{\rm BS}) = \frac{1}{\epsilon}$ with the corresponding residue being \cite{Economou2006}
\begin{equation}
	\text{ Res } \left[ {G} ( z ) , z = \omega_{\rm BS}\right] =   \frac{ 1 }{ \braket{0| {G}_B ^2 ( \omega_{\rm BS})|0 } }\,\ket{\psi ( \omega_{\rm BS}) } \bra{\psi ( \omega_{\rm BS})},
\end{equation}
The BS is non-degenerate since the degeneracy is calculated as $\text{Tr}\, \{\!\!\text{ Res } \left[ {G} ( z ) , z = \omega_{\rm BS}\right] \}= 1$ \cite{Economou2006}. The normalized BS wavefunction is given by ${\cal N}\ket{ \psi ( \omega_{\rm BS}) }$ with ${\cal N}=\braket{0|{G}_B^2 ( \omega_{\rm BS}) |0 }^{-1/2}$.

Applying the Lippmann-Schwinger equation, an unbounded eigenstate of energy $\omega$ is given by
\begin{equation}
	\ket{\psi^\pm _k } = \ket{ k ( \omega ) } + {G}^\pm ( \omega ) V_{\rm imp} \ket{k ( \omega ) }.
\end{equation}
This gives
\begin{equation}
	\ket{\psi^\pm _k } = \ket{ k ( \omega ) } + \lim_{\delta \rightarrow 0} \frac{1}{f ( \omega \pm i \delta )  }\braket{0|k ( \omega \pm i \delta ) } \ket{\psi ( \omega \pm i \delta ) },
\end{equation}
for each $\omega$ such that $ \gamma_0(\omega) \neq \frac{1}{\epsilon}$ [namely $f(\omega)\neq0$]. If instead $\gamma_0(\omega) = \frac{1}{\epsilon}$, then $\ket{\psi^\pm _k }= \ket{ k ( \omega ) } $.

\section{Unbound dressed states}\label{app-LS}

Similarly to the previous section, based on the Lippmann–Schwinger equation, an unbound dressed state of energy $\omega$ fulfils 
\begin{equation}
	\ket{\psi_k } = \ket{ k ( \omega ) } + {G}^+ ( \omega ) V_{\rm imp} \ket{k ( \omega) }\,.
\end{equation}
Using \eqs\eqref{Hd}, \eqref{psiz} and \eqref{fz} (see main text) then yields \eq\eqref{psik} in the main text.

In our case, the modes that belong to the bands are the modes of the bare bath, and this equation simply becomes
\begin{equation}
	\label{eq:lipswin}
	\ket{\psi_k } = \ket{k ( \omega ) } +  \frac{\braket{0 | k ( \omega )}}{ F^+ ( \omega )} \ket{\Psi( \omega )} .
\end{equation}
To work out a more explicit expression for the unbound dressed states, we recall the identities
\begin{align}
	\lim_{y \rightarrow 0} \frac{1}{x \pm i y}  = {\cal P}\frac{1}{x} \mp i \pi \delta ( x ) \,,\,\,\,\delta ( g ( x ) ) = \sum_{i }\frac{ \delta ( x - x_i )}{ | g' ( x_i ) | } \,\,\,{\rm with}\,\, g( x_i ) = 0 \,\, 
\end{align}
and expand \eq\eqref{eq:lipswin} as
\begin{equation}
	\ket{\psi_k } = \ket{k ( \omega ) } +  \braket{0 | k ( \omega )} {\cal P}\tfrac{1}{ F ( \omega )} \ket{\Psi( \omega)} - i \pi  \sum_{ i } \frac{\braket{0 | k ( \omega_i )} \delta( \omega - \omega_i )}{ \frac{1}{g^2} +  \braket{ 0|{G}^2_B ( \omega_i )|0}}  \,\,\ket{\Psi( \omega_i)} ,
\end{equation}
with $F(\omega_{i })=0$. If $F(\omega) \neq 0$, only the second term (featuring the principal part) adds to $ \ket{k ( \omega ) }$. If instead $F(\omega) = 0$ (namely $\omega$ coincides with some $\omega_{i}$) then the only possible corrrection to $ \ket{k ( \omega ) }$ can come from the third term. This is yet zero since 
\begin{equation}
\braket{0 | k ( \omega_i )}=0\label{van}\,.
\end{equation}
To show this, we note that the bath resolvent on the real axis can be expanded as
\begin{equation}
	G _B ( \omega ^\pm) = \int {\rm d } \omega'\, { \cal P }\,\frac{ \rho_B ( \omega') \ket{k ( \omega' ) } \bra{ k ( \omega' ) }}{ \omega - \omega'} \mp i \pi \rho_B ( \omega ) \ket{ k ( \omega ) } \bra{ k ( \omega )},
\end{equation}
where the density of state $ \rho_B ( \omega ) $ is $ \frac{\partial k}{\partial \omega} $.
Taking the expectation value of both sides on state $\ket{0}$, we see that ${\rm Im}\,\langle 0|G _B ( \omega ^\pm)|0\rangle=0$ if and only if \eqref{van} holds.
Then if there exists a solution of $F ( \omega ) = 0 $ in the continuum then the associated eigenstate of the same energy of the bare bath is unaffected by the atom, entailing $\ket{\psi_k } = \ket{k ( \omega_i ) }$.  

Thus \eqs\eqref{psik} in the main text are proven.
\section{Two atoms}
%

Based on \eqs\eqref{G2at} and \eqref{F2} in the main text, the two-atom BS energy $\omega_{\rm BS}$ fulfils the two-atom pole equation ${\det {\bf F}} ( \omega_{\rm BS} ) = 0$. The solution is given implicitly by
\begin{equation}
	\omega_{\rm BS} ^\pm = {\tilde \omega}_0(\omega_{\rm BS} ^\pm) \pm \delta(\omega_{\rm BS} ^\pm)
\end{equation}
with ${\tilde \omega}_0(z)$ and $\delta(z)$ depending on $z$ according to
\begin{equation}
	{\tilde \omega}_0^\pm  = \omega_0 + \tfrac{g^2}{2}   \left(\braket{  {G}_B(z)  }_{x_1} { +}  \braket{ {G}_B(z) }_{x_2}\right)\,,\,\,	\delta(z)  =  \sqrt{g^4 F_{12}   F_{21}+{\cal A}^2}\,,
\end{equation}
with the asymmetry coefficient ${\cal A}$ defined as
\begin{equation}
{\cal A}(z)  = \tfrac{g^2}{2} \left(F_{22}(z)-  F_{11}(z)  \right)\,.
\end{equation} 

The residues of $G(z)$ on the two poles $\omega_{\rm BS}^\pm$ can be calculated directly, yielding (we use a vector-matrix formalism)
\begin{equation}\label{Res2}
	\ket{\Psi_{\rm BS}^\pm}\! \bra{\Psi_{\rm BS}^\pm} = \Res{\omega_{\rm BS}}\left. \left[ {G} ( z ) \right] = \frac{g^2}{\beta} \left[ \begin{pmatrix}
		\ket{\Psi_{1}} & \ket{\Psi_2}
	\end{pmatrix} \begin{pmatrix}
		F_{11} & -F_{12}\\
		-F_{21} & F_{22} 
	\end{pmatrix} \begin{pmatrix}
		\bra{\Psi_{1}} \\ \bra{\Psi_2}
	\end{pmatrix}	 \right]\right|_{z=\omega_{\rm BS}^\pm}
\end{equation}
with
\begin{equation}
\beta(z) =g^2 \left[F_{11}  \braket{ \Psi_2 | \Psi_2 } + F_{22} \braket{ \Psi_1 | \Psi_1} -   F_{12} \braket{ x_2 | {G}_B^2 | x_1} - F_{21} \braket{ x_1 | {G}_B^2 | x_2} \right]\,
\end{equation}
[the dependence on $z$ is left implicit in the expressions for $G_B(z)$, $F_{ij}(z)$ and $\Psi_i(z)$].

In the weak-coupling regime, we can make the following approximations
\begin{align}
	{\tilde \omega}_0(\omega_{\rm BS}^\pm) &\simeq {\tilde \omega}_0(\omega_0)=\omega_0 + \tfrac{g^2}{2}  \left[ \braket{  {G}_B ( {\omega}_0  ) }_{x_1}  +  \braket{ {G}_B ( {\omega}_0  ) }_{x_2} \right]\,, \\
	F_{ij}(\omega_{\rm BS}^\pm)  & \simeq F_{ij} ( \omega_0 )  \pm \tfrac{\delta (  {\omega}_0 )}{g^2} \delta_{ij}\,,\,\,\,\mathcal{A}(\omega_{\rm BS}^\pm)  \simeq\mathcal{A}(\omega_0)=\mathcal{A}_0\,,\,\,\,\ket{\Psi_i(\omega_{\rm BS}^\pm) }  \simeq\ket{\Psi_i ( { \omega}_0 )}=\ket{\Psi_i} ,\\
	\delta(\omega_{\rm BS}^\pm)  &\simeq  \delta(\omega_0)=\delta_0\,,\,\,\,\beta(\omega_{\rm BS}^\pm)  \simeq\beta^\pm=  {\cal A}_0 \left( \braket{ \Psi_1 | \Psi_1}  -  \braket{ \Psi_2 | \Psi_2} \right)\pm \delta_0 ( \braket{ \Psi_2 | \Psi_2} + \braket{ \Psi_1 | \Psi_1} )  \,,
\end{align}
hence \eqref{Res2} reduces to
\begin{equation}
	\ket{\Psi_{\rm BS}^\pm} \!\bra{\Psi_{\rm BS}^\pm} = \frac{1}{\beta^\pm } \left[ ({\cal A}_0 \pm \delta_0) \ket{\Psi_{1}}\bra{\Psi_{1} } + (-{\cal A}_0 \pm \delta_0 )  \ket{\Psi_2 }\bra{\Psi_2 } - g^2 F_{12} \ket{\Psi_{1}}\bra{\Psi_{2} } - g^2 F_{21} \ket{\Psi_{2}}\bra{\Psi_{1} }	 \right]\,.
\end{equation}
The full Hamiltonian can be spectrally decomposed as
\begin{equation}\label{H-SD}
	H = \sum_{i=\pm} \omega_{\rm BS} ^i \ket{\Psi_{\rm BS}^i}\! \bra{\Psi_{\rm BS}^i} + {\rm unbound\,\,dressed\,\,states}\,.
\end{equation}
For $\omega_{0}$ out of the photonic continuum and $g$ sufficiently small, states $\ket{e_1}$ and $\ket{e_2}$ will negligibly overlap unbound states (as these are far-detuned). Hence, the atomic dynamics is in this case fully captured by the BSs contribution in \eq\eqref{H-SD}. Accordingly, the effective atomic Hamiltonian is thus given by \eq\eqref{Heff1} in the main text, where $\omega_{\rm BS}^{(i)}  = \omega_0 + g^2 \braket{ G_B ( \omega_{\rm BS}^{(i)} ) }_{x_i}.$
\section{Many-atom effective Hamiltonian}
Recall from \eqs\eqref{G2at} and \eqref{F2} the many atom BS energy equation   
\begin{align}\label{pole}
	\det \bm{F}(z)=\tfrac{1}{g^2}\det[z-\omega_0-g^2\bm{\gamma^B}(z)]=0\,.	
\end{align}
This is a non-linear equation which yields $M$ (possibly coinciding) solutions $\{\omega_{\rm BS}^{(i)}\}_{i=1}^M$. Correspondingly, the bound states can be calculated as the residues of $G(z)$ on the poles $\omega_{\rm BS}^{(i)}$, which can be expressed
\begin{equation}\label{Res3}
	\ket{\Psi_{\rm BS}^{(i)}}\! \bra{\Psi_{\rm BS}^{(i)}} = \Res{\omega_{\rm BS^{(i)}}}\left[ {G} ( z ) \right] =\sum_{jk}
	\left\{\Res{\omega_{\rm BS^{(i)}}}[\bm{F}^{-1}(z)]\right\}_{jk} 
	\ket{\Psi_{j}(\omega_{\rm BS}^{(i)})}\bra{\Psi_{k}(\omega_{\rm BS}^{(i)})}
\end{equation}
In the weak-coupling regime, i.e. for $g\ll\Delta$ where $\Delta$ is the detuning of $\omega_0$ from the continuum of frequencies (\eg\, energy bands)  of the lattice, equation~\eqref{pole}, up to second order in $g/\Delta$, can be approximated by $\det[z-\omega_0-g^2\bm{\gamma^B}(\omega_0)]=0$. Thus, decomposing $\bm{\gamma}^B(\omega_0)=\sum_i\gamma_i|\gamma_i)(\gamma_i|$ in terms of its eigenvalues $\gamma_i$ and eigenvectors $|\gamma_i)$ ($\gamma^B(z)$ is Hermitian, for  $z\in\mathbb R$) yields
\begin{align}
	\omega_{\rm BS}^{(i)}&\simeq\omega_0+g^2\gamma_i\\
	\Res{\omega_{\rm BS^{(i)}}}\left[ \bm{F}^{-1} ( z ) \right]&\simeq g^2|\gamma_i)(\gamma_i|\,.
\end{align}
%
As for the two-atom case, in the weak coupling regime, the atomic dynamics is described by the BSs contribution of the full Hamiltonian analogue of \eq\eqref{H-SD}. Hence, the effective Hamiltonian is given by
\begin{align}
	H_{\rm eff} &\simeq \sum_{i} \omega_{\rm BS} ^{(i)} \ket{\Psi_{\rm BS}^{(i)}}\! \bra{\Psi_{\rm BS}^{(i)}}\simeq g^2\sum_{jk}\left\{\sum_i(\omega_0+g^2\gamma_i) |\gamma_i)(\gamma_i|\right\}_{jk}\ket{\Psi_j}\bra{\Psi_k}\nonumber\\&=g^2\sum_{jk}\{\omega_0\one_M+g^2\bm{\gamma}^B(\omega_0)\}_{jk}\ket{\Psi_j}\bra{\Psi_k}=g^2\sum_{jk}\left[\omega_0\delta_{jk}+g^2\braket{x_j|G_B(\omega_0)|x_k}\right]\ket{\Psi_j}\bra{\Psi_k}\,,\label{HeffM}
\end{align}
which, upon renormalisation of the states $\ket{\Psi_i}$, coincides with \eq\eqref{eq:HeffM} in the main text.

%

\twocolumngrid

\bibliographystyle{iopart-num}
\bibliography{WQEDdefects}

\end{document}